# Quantifying the Cognitive Extent of Science

*Staša Milojević[a]*

[a] *School of Informatics and Computing, Indiana University Bloomington 47405-1901, United States*

*E-mail addresses:* smilojev@indiana.edu

## Abstract

While the modern science is characterized by an exponential growth in scientific literature, the increase in publication volume clearly does not reflect the expansion of the cognitive boundaries of science. Nevertheless, most of the metrics for assessing the vitality of science or for making funding and policy decisions are based on productivity. Similarly, the increasing level of knowledge production by large science teams, whose results often enjoy greater visibility, does not necessarily mean that "big science" leads to cognitive expansion. Here we present a novel, big-data method to quantify the extents of cognitive domains of different bodies of scientific literature independently from publication volume, and apply it to 20 million articles published over 60-130 years in physics, astronomy, and biomedicine. The method is based on the lexical diversity of titles of fixed quotas of research articles. Owing to large size of quotas, the method overcomes the inherent stochasticity of article titles to achieve <1% precision. We show that the periods of cognitive growth do not necessarily coincide with the trends in publication volume. Furthermore, we show that the articles produced by larger teams cover significantly smaller cognitive territory than (the same quota of) articles from smaller teams. Our findings provide a new perspective on the role of small teams and individual researchers in expanding the cognitive boundaries of science. The proposed method of quantifying the extent of the cognitive territory can also be applied to study many other aspects of 'science of science.'

## 1   Introduction

Modern, post-World War II science is characterized by two prominent features: an exponential growth in the output of scientific literature (Bornmann & Mutz, 2015; Price, 1961, 1963; Tabah, 1999; van Raan, 2000), and the increase in the level and extent of collaboration among researchers (Bordons & Gomez, 2000; Guimerà, Uzzi, Spiro, & Amaral, 2005; Jones, Wuchty, & Uzzi, 2008; Milojević, 2014; Wuchty, Jones, & Uzzi, 2007). The growth of literature is also reflected in the increase of publication venues (Kaiser, 2012; Kochen, 1974; Menard, 1971; Pautasso, 2012) and the rising number of researchers (Stephan, 2012; Teitelbaum, 2014; Xie & Killewald, 2012). Specifically, the volume of articles produced annually has been estimated to grow between ~5% (Larsen & von Ins, 2010) and 8 to 9% (Bornmann & Mutz, 2015) a year, i.e., the scientific output is doubling every 15 to 9 years, respectively.

The rise of collaborative practices that only a few decades ago characterized a handful of disciplines is now prevalent in most scientific fields (Shrum, Genuth, & Chompalov, 2007; Wagner, 2008). Research teams are not only becoming more common, they are also increasing in size, as evidenced by the sheer drop in the fraction of single-authored papers (Wuchty et al., 2007), and the shift of team-size distributions from small teams (<10 authors) to large, power-law distributed teams (10-1000 authors) (Milojević, 2014). The changes in the character of knowledge production in the period after World War II were so profound in some fields that they warranted the introduction of the term "big science," as opposed to "little science," to describe an emerging class of practices used in tackling modern scientific problems (Galison, 1997; Price, 1963; Weinberg, 1961). Five decades after its introduction, "big science," and large-team efforts in general, appear to be trumping "little science" on many fronts. Large teams are credited with higher impact results (Börner et al., 2010; Tabah, 1999) and with more innovative research (Uzzi, Mukherjee, Stringer, & Jones, 2013).

The growth of scientific literature suggests that the extent of the cognitive territory of science must be expanding, and, consequently, productivity-based metrics are often used in connection with the funding of science and in making science policy decisions (Kaiser, 2012; Stephan, 2012; Teitelbaum, 2014). Nevertheless, most researchers and policy makers are acutely aware that the increased output does not necessarily correspond to the expansion of the cognitive domains of science. Cole & Cole (1972), for example, differentiated between the mere productivity and contributions to scientific progress, examining the relative contributions to the latter by "elite" and "non-elite" scientists. Furthermore, it is not obvious whether large scientific teams, despite their evident successes, have a principal role in expanding or even maintain the cognitive boundaries of science. In order to properly quantify the cognitive domains of scientific fields, research areas or of specific groups of scientists, and their change through time, one must explore metrics other than those based on productivity.

Here we propose a new, big-data metric of science that is more appropriate for exploring its cognitive content than the productivity-based measures. The new metric is based on the analysis of the text appearing in the titles of a large number of scientific articles. The use of text in the quantitative study of science has a long and rich history. Textual bibliographic information forms the basis of many studies that aim to reveal the cognitive structure of science through, for example, co-word analysis (Braam, Moed, & van Raan, 1991a, 1991b; Callon, Courtial, & Laville, 1991; Courtial & Law, 1989; Leydesdorff, 1989; Milojević, Sugimoto, Yan, & Ding, 2011; Noyons & van Raan, 1998; Rip & Courtial, 1984; van Raan & Tijssen, 1993; Whittaker, Courtial, & Law, 1989). Other studies have applied lexical analysis methods to identify emerging topics and research fronts in science (e.g., Zitt, 1991; Zitt & Bassecoulard, 1994). More recently, a number of studies used text analyses to address the questions of novelty and creativity in science. For example, Tang & Hu (2013), used the corpus of "keywords" to examine the linkage between international collaboration and knowledge spillover. Franzoni (2010) successfully used phrases from titles and abstracts to measure creativity of physicists, while Foster et al. (2015) used MEDLINE to study research strategies in biomedical chemistry, tracking chemicals and chemical relationships extracted from abstracts. Recent work (e.g., Milojević, 2012; Milojević et al., 2011) has demonstrated that useful information regarding the cognitive content of science is contained even in short text of article titles. Title text, when used on large scale, has the advantage of being more distilled compared to abstracts and is richer and less constrained than the keywords.

The approach for quantifying the extent of cognitive domains of scientific fields presented in this work is based on the concept of *lexical diversity* (e.g., Jarvis, 2013; Malvern, Richards, Chipere, & Durán, 2004), but operates on the level of concepts (phrases) rather than words, and also overcomes critical limitations of the classical methods to measure lexical diversity.

In addition to presenting the new method, we also apply it to elucidate the roles of different modes of knowledge production, specifically, to explore whether big-team science has rendered the efforts of individual authors and small teams obsolete. Rather than using standard, productivity-based metrics, here we answer this question by studying how the extents of cognitive territories of different modes of knowledge production compare.

*The proposed method*

The method for quantifying the cognitive extent of scientific fields proposed in this paper exploits the fact that the titles of journal articles contain information that reveals their cognitive content (Bazerman, 1988; Callon, Courtial, Turner, & Bauin, 1983; Kuhn, Perc, & Helbing, 2014; Leydesdorff, 1989). Specifically, we propose that the number count of different *unique phrases* appearing in the titles of statistically large unit quotas of scientific literature (thousands of articles) reflects the extent of the cognitive territory covered in those bodies of literature, and can therefore be used to measure it. This approach is similar to one taken to determine the lexical diversity or richness of a text (Jarvis, 2013; Malvern et al., 2004), but operates at a level of phrases, rather than single words. Phrases consist of one or more words, derived by processing and parsing article titles using some simple rules and exclusion dictionaries. Because the titles of scientific articles, as opposed to full text of articles or even the abstracts, are distilled conceptually, the phrases extracted in this way very closely correspond to scientific concepts, which form the unit elements of the cognitive territory, the extent of which we wish to measure. In other words, we use title phrases as a proxy for scientific (cognitive) concepts. The use of multi-word phrases, as opposed to single words, is essential, as it allows one to capture new concepts that arise from the interaction of the existing ones. For example, *mass-energy equivalence* would be counted as a separate phrase from either *mass* or *energy*. A single title will usually contain several concepts. The method does not equate the abundance of title concepts with knowledge, an inherently abstract concept, but represents a step towards the quantification of some of its more tangible aspects. As such, it represents a departure from typical science metrics that focus on the volume of literature.

The proposed method may appear limited because any particular article title obviously represents only one of many possibilities that the authors had at their disposal when choosing the title, and will therefore reflect the underlying cognitive extent with some degree of stochasticity. Furthermore, the title will contain only a limited number of concepts covered in the work, presumably the most important ones. The linguistic style will also vary from case to case, with some authors aiming for catchy or amusing titles (which nevertheless usually contain cognitive concepts). However, as we will show, the confounding effects of an incomplete and imperfect sampling are erased because the proposed measure is based on a statistically large number of article titles. The statistical nature of the method also addresses the fact that not every title concept carries the same significance, because the central limit theorem ensures that the total count of concepts (a mix of phrases with a varying degree of significance) will be proportional to the overall underlying cognitive content that these concepts sample[1]. The method currently does not collate synonymous concepts. This means that the cognitive measure will be overestimated by some, probably small factor. However, this will not affect the *relative* measurements and trends, which are at focus here, because there is little reason to expect that the fraction of concepts that are synonymous would greatly vary with time or with team size.

---

[1] One may argue that an ever increasing number of objects of study in some disciplines may artificially inflate the number of unique concepts. For example, of the fields studied here that may be the case with astronomy, where titles pertain to specific galaxies or stars, identified by their catalog numbers. To reduce this potential bias we exclude from counting all phrases consisting of numbers.

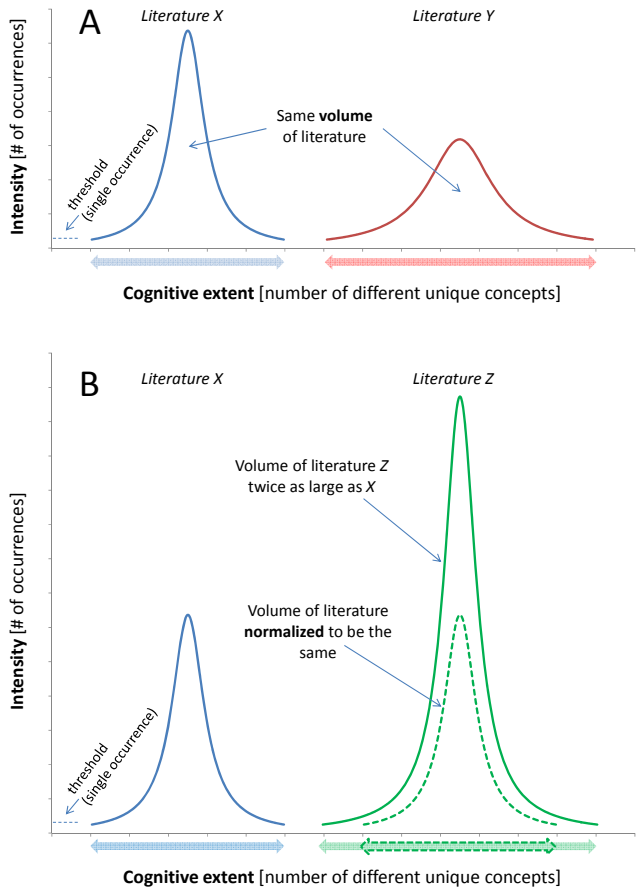

**Fig. 1. Schema of the method for quantifying the cognitive extent of scientific literature.** The cognitive extent of some body of literature will be proportional to the number of unique concepts appearing in article titles assuming that the volume (total number of concepts) is the same in each body of literature. Panel A shows a case where literature *Y* has a greater cognitive extent than literature *X* (wider base). Panel B shows that without using the same unit quota in which to measure the concepts, the literature *Z* would appear to have a greater extent than *X*.

One of the limitations of classical measures of lexical diversity is that the number of unique words, even if normalized by the length of the text (the TTR, or type-per-token measure), will strongly depend on the length of the text (e.g., Koizumi, 2012; McCarthy & Jarvis, 2007). Techniques have been developed to provide corrections for this effect (e.g., the *vocd* method (McKee, Malvern, & Richards, 2000)), but these corrections can be uncertain and nevertheless depend on the length of the text (McCarthy & Jarvis, 2007). We remove this dependency entirely by using the *unit quotas*. The necessity of applying the lexical diversity measure to *unit quotas* of literature is illustrated schematically in Figure 1. Literature Y has greater lexical diversity (hence, greater cognitive extent) than literature X only if the number of unique concepts in Y is larger than in X while the volume of literature X and Y is the same (Figure 1A). The measure of the cognitive extents of the bodies of literature must be applied to the same volumes of literature (unit quotas), because the probability of phrase occurrence reaching a counting threshold (an occurrence) will simply scale with the volume of text (in this case, the titles). Figure 1B shows the case where literature Z appears to have a larger cognitive extent than X, but after its volume has been scaled down to that of X (e.g., by sampling), the extents are revealed to be the same.

We adopt the unit quota of scientific literature that corresponds to approximately 3,000 articles. Because the lengths of article titles have been changing over the decades, the unit quota is actually defined as a fixed number of title phrases (10,000), rather than a fixed number of titles. The choice for the size of the unit quota is discussed in Section 2. The principal limitation of the method is that the quota cannot be smaller than ~1,000 phrases, because the number of unique phrases in small quotas does not change linearly with lexical diversity (what we call saturation, and discuss more in Section 2), which means that it cannot be applied to measure cognitive extents of small bodies of literature (e.g., works of individual researchers). Additionally, the titles need to be in one language.

Literature (in our case, article titles) can be grouped into unit quotas according to any number of criteria. In this paper we group articles by publication periods, in order to study cognitive evolution, and by team sizes, to study the relative cognitive extents of their scientific output.

## 2 Data and analysis

*Data*

In this study we are interested in quantifying the cognitive extents of several broad disciplines (fields). The scope of the discipline will be defined by the journal articles that are included in the analysis. We therefore make every attempt to include representative journals for a given field. The fields studied are: physics (440,000 articles), astronomy (160,000 articles), and biomedicine (19,600,000 articles), with articles published over extensive time periods of 117, 125, and 67 years, respectively. The physics dataset contains only American journals, while astronomy and biomedicine datasets are international.

For physics, we use all journals published by the American Physics Society (APS). They cover broad areas of physics and are one of the principal venues for publication of research in physics (Kuhn et al., 2014; Perc, 2013). APS journals include: Physical Review (PR) Series I (1893-1913), II (1913-1970) and III (since 1970). Series III is split into sub-journals (A, B, C, D, E, STAB, and STPER). We also include Physical Review Letters (PRL, since 1958). Bibliographical data for these journals, covering the time period 1893-2009, were obtained from APS website in November 2013. We remove items whose titles indicate that they are comments, replies or errata on other papers, as well as publisher notes. The final physics dataset consists of 441,623 articles.

For astronomy, we select articles published in four main astronomy journals (Henneken et al., 2007): Monthly Notices of Royal Astronomical Society (1827), Astrophysical Journal (1895), Astronomical Journal (1849) and Astronomy & Astrophysics (1969). These journals publish the majority of all articles in astronomy. Years in which these journals began publishing are given in parentheses. For the analysis in this paper we focus on the time period 1886-2010, during which continuous data are available for at least one journal. Astrophysical Journal includes Supplement Series and Letters. Data for articles prior to 1955 (prior to 1960 for Astronomical Journal) were retrieved from the NASA Astrophysical Data System in February 2014. Data for subsequent years have been obtained from the Thomson Reuters Web of Science (WoS) database in October 2012 (except for Astronomical Journal, in July 2012). From ADS we accept items classified as refereed, and from WoS we accept items classified as "article" or "letter." We remove duplicate entries that arise from database errors and also items that are derivative in nature (comments or replies to other papers and errata). The final astronomy dataset consists of 155,429 articles.

Bibliographical data for biomedical literature comes from PubMed, retrieved from the Scholarly Database in April 2013 (Light, Polley, & Börner, 2013; Rowe, Ambre, Burgoon, Ke, & Börner, 2009). While PubMed contains entries starting from 1809, we use data for the period 1946-2012, for which the data appear to be largely complete. After removing derivative titles the PubMed dataset contains 19,573,404 articles.

For each paper we establish the size of the team that produced it as the number of (co)authors that are listed on the paper. Teams can also be defined in different ways (e.g., a group of people collaborating over some period of time), but the current definition is both easier to operationalize and is also more useful to understand the correlation between the knowledge producers and the knowledge they produce.

*Processing of titles and parsing into phrases*

Before phrases can be extracted from them, the titles were processed in the following way: Periods, colons, semicolons, dashes, and commas were replaced with a code specifying phrase separation. All punctuation symbols were removed, except for ampersands, which were replaced with "and." Leading and trailing spaces were removed, and multiple spaces were replaced with a single space. Cases were ignored. Numbers and single letters were removed. Singular and plural forms of nouns were collated.

Identification of title phrases is performed automatically. A phrase is defined as the longest string of words separated by general word(s), or a phrase delimiter. A phrase delimiter is one of the symbols used to designate the subtitle (period, colon, semicolon, dash) or a list of phrases (comma). General words are words that appear in the title but do not specify a cognitive concept. The list of general words is constructed by combining a list of 581 common English words (such as *about, down, how, none, since, using*, etc.) supplied with the text analysis software Wordstat, with an additional list of less common words that nevertheless do not specify a scientific concept. Determining if a word is specific or non-specific requires manual inspection. It is impractical to inspect every title word. Instead we manually classify only the most commonly found words. Choosing words to classify by straight word frequency would bias the classification towards the common words occurring in recent years in which the volume of article is largest. Instead, we define the most common words to be those that appear in largest number of publication years. Specifically, we classify all words that appear in at least 50 different publication years (there are 1,549 such words in physics and 878 in astronomy), observing the context in which they are found. For biomedicine, the number of words that appear in all years of the coverage is excessive, so we inspect 1,000 of the most frequent words weighted by the inverse of the volume of articles in that year, again in order to obtain a list that reflects the entire period of study in a balanced way. In addition, because many of the non-specific words are past tenses of verbs (e.g., *related, proposed, based, derived*), we also inspect all words ending in "ed" that appear in at least thirty years for physics and astronomy, and among the top 2,000 words for biomedicine. For physics 168 words are classified as non-specific, the most common being: *method, determination, study, note, property, measurement*, and *phenomenon*. For astronomy 253 words are non-specific. The most commonly appearing ones are: *determination, theory, observed, method, list, note, comparison*, and *study*. Inspection of frequent words in biomedical literature yielded 110 non-specific words, such as *study, effect, analysis, induced, report, role, change, evaluation*, etc. Identification of non-specific words among less common words brings diminishing returns. Namely, even if we use only half of the identified non-specific words to parse the phrases, we still obtain the same trends (both large-scale and fine-scale) in the measure of cognitive extent. More sophisticated algorithms for the identification of concepts can be tested in the future, but the current methods suffice for the intended purpose.

It is not uncommon, especially in titles of recent articles, to encounter compound phrases consisting of a large number of adjectives (e.g., *high resolution energy filtered scanning tunneling microscopy*). Whether some phrase represents a separate concept, a cognitive unit, depends on the desired level of granularity. In this study we aim for an intermediate level. We establish that this is best achieved by limiting the number of words that can make a phrase to three words. Note that we do not discard phrases that have more than three words, but rather take the last three words to represent a cognitive unit (e.g., *high resolution energy filtered scanning tunneling microscop*y and *low temperature scanning tunneling microscopy* would both be considered the same phrase *scanning tunneling microscopy*.) As in the case of the extensiveness of the list of general words, the results are very robust under modified assumptions. For example, allowing up to four words to define a phrase leads to the change in the cognitive extent of only 1-2%, with no change in the details of trends.

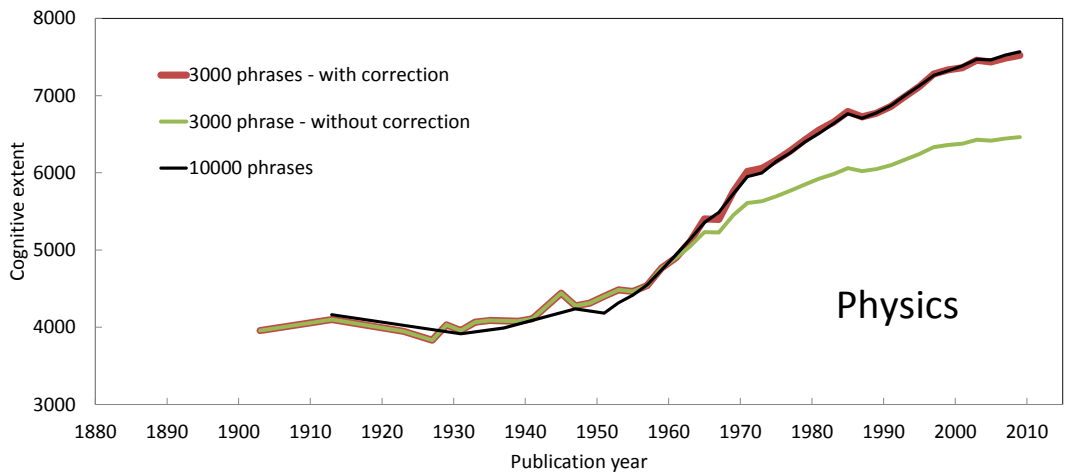

**Fig. 2. The effect of the size of a unit quota on the measure of the cognitive extent**. Using smaller unit quotas (3000 phrases, green line) gives finer temporal resolution at early times, where publication volume is small, but is subject to a saturation, which is reflected in its falling below the trend obtained with a larger quota (10000 phrases, black line). Quotas larger than 10,000 behave similarly to 10,000, i.e., the quota of 10,000 is not affected by saturation and we adopt it as a reference quota. For physics and astronomy we perform calculations with a smaller quota, and then apply an empirical linear correction to make the measurements agree with those that would have been obtained with a larger quota (red line). The physics dataset is shown. Similar behavior is seen for astronomy. In biomedicine the volume of articles is sufficiently large to only use the full quota.

*Choice of unit quotas*

We measure cognitive extent as the number of unique phrases in a given amount of title text, a unit quota, set to contain 10,000 phrases. Using smaller quotas (e.g., 3,000 phrases) has the advantage that at early times, when the number of papers published annually was smaller, the cognitive extent can be measured with finer temporal resolution. However, a smaller quota also means that the measure would saturate when the number of unique concepts starts to approach the size of the quota (Figure 2), i.e., the number of unique concepts will no longer depend linearly on the underlying lexical diversity. We have empirically established that for article titles the saturation is no longer present if the quota is larger than 10,000 phrases, which is why we choose it as the reference size of unit quota. However, we still wish to exploit the advantages of a smaller quota. We thus find an empirical scaling that correlates the number of unique phrases in a smaller quota (3000 phrases, $N_u(3000)$) to the reference quota ($N_u(10000)$), i.e., corrects for the non-linearity when lexical diversity is high. The scaling factor for physics titles is:

$$\frac{N_u(10000)}{N_u(3000)} = 6.65 \times 10^{-4} N_u(3000) + 1.19, \text{ for } N_u(3000) > 1970$$

and for astronomy:

$$\frac{N_u(10000)}{N_u(3000)} = 5.71 \times 10^{-4} N_u(3000) + 1.40, \text{ for } N_u(3000) > 1926$$

Below these values of $N_u(3000)$ the saturation is not present and the scaling factor is a constant 2.5 for both fields. The high accuracy of these scaling relations can be assessed from Figure 2. For biomedicine, the volume of articles is 40-90 times greater than in physics and astronomy, so we are able to measure the cognitive extent using the reference quota of 10,000 phrases directly, without losing temporal resolution.

During the earlier years of physics and astronomy, where the publication volume is low, the temporal resolution obtained with a unit quota of 3,000 phrases is 10 years or less. Afterwards, the volume of articles is sufficient to measure cognitive contents from multiple full quotas even within the single year. The standard deviation of these contemporaneous measurements is only 0.5 to 1.5% in recent times, implying high intrinsic precision of the method. Averaged over 2-yr periods, the random error is less than 0.2%. Note that it is important to randomize the order of articles in a given year, so that each unit quota samples all of the different journals present in the dataset.

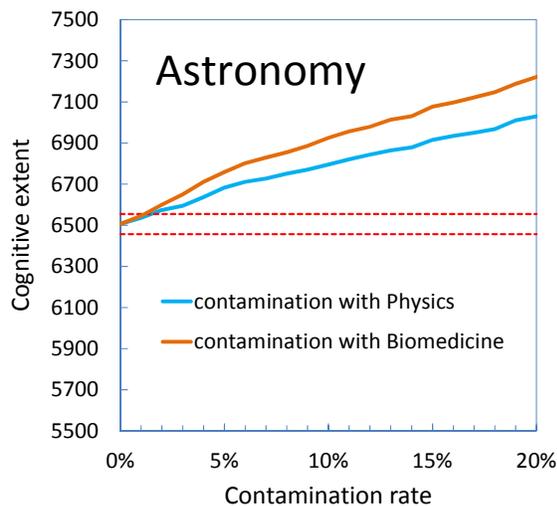

**Fig. 3. The sensitivity of the method assessed through a simulated contamination of articles from astronomy with those from physics or biomedicine**. Test shows that even 1-2% addition of phrases from another field results in a measure that deviates from the range expected from statistical fluctuation (dashed lines), as determined by jackknife resampling.

*Validation and testing*

The measure of the cognitive extent could be affected by publishing practices of specific journals. In the case of astronomy we have several journals that are all supposed to cover the entire field. We have verified that we obtain similar measures of the cognitive extent (within 5%) if we use any of these journals alone. This suggests that the measure is not very sensitive to the exact selection of input data. Nevertheless, it is advisable, as we do, to use at least several journals to ensure that any specificities of one do not affect the final result.

We test the sensitivity of the method by taking a single quota of articles in astronomy, and then replacing a fraction of articles with articles from physics or from biomedicine, and observing how quickly the measurement of the cognitive extent changes as a result of "contamination." We first estimate the level of uncertainty of the non-contaminated astronomy quota measurements by performing jackknife sampling, a version of bootstrap resampling. We draw, at random, one half of the articles from two successive quotas and calculate the extent of the cognitive content. We perform ten random drawings and find the standard deviation of the values obtained in each run. The range corresponding to ±1 standard deviation is shown Figure 3 as dashed lines. It corresponds to 0.8%, in line with the estimates of the scatter from multiple contemporaneous quotas. The result of adding contaminating articles is visible in Figure 3 as the departure of blue and red curves from the middle of the dashed band. It shows that it is sufficient to have 1-2% of articles from another field (even a related field, such as physics) in order to perturb the resulting measure beyond its statistical fluctuation. This sensitivity allowed us to identify an error in WoS data. Namely, we have noticed a spike in the measure of cognitive extent in astronomy for the year 1963. Inspecting the titles of articles from that year we have found that it included 46 articles with topics from medicine. We have tracked these articles to journal Lancet. They were incorrectly entered in WoS as being published in the Astrophysical Journal. We have removed them from the final sample.

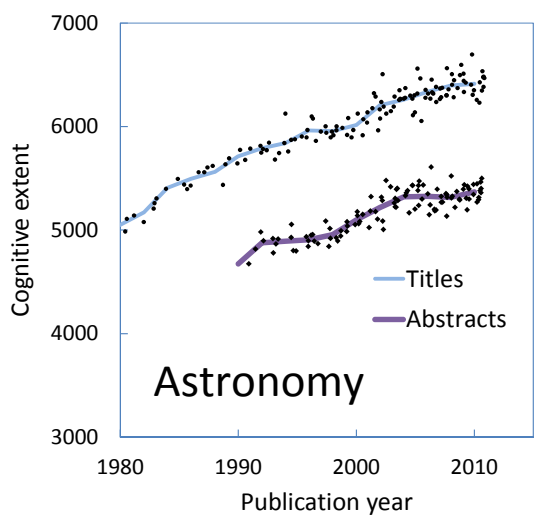

**Fig. 4. Comparison of the trends of cognitive extent as derived from abstracts and titles**. Abstracts produce similar trends as the titles (trends offset for clarity), with similar level of precision (scatter around the mean trend). Abstracts are available in our dataset only for astronomy articles published since 1991.

Finally, we test the degree to which the results would differ if they were based on the analysis of abstracts rather than the titles. Abstracts, typically containing 20 times as much text as the titles are naturally richer sources of cognitive content than just the titles. However, the concepts that they contain will be less distilled than the ones in the titles, so they do not necessarily carry the proportionally greater amount of information. Furthermore, abstracts are not as readily and extensively available in bibliographic databases as the titles. Of the datasets that we had in this study, only the WoS data (astronomy) included abstracts, and only for articles published since 1991. We analyzed these abstracts in the same way as the titles, but using a larger unit quota (50,000 abstract phrases). The resulting two-decade trend of cognitive extent is shown in Figure 4, alongside the trend based on titles (offset for clarity). The trends have common salient features. The scatter of points around the average trend is similar for abstracts and titles, suggesting comparable intrinsic precision. Testing the method against the full text of articles (the large-scale availability of which is even more problematic than that of abstracts) is beyond the scope of this article. Even more than in the case of the abstracts, the greater amount of text would not necessarily present an advantage, because of the overabundance of phrases that may not be particularly important for the determination of the cognitive content. The use of keywords is not recommended because the pool of keywords is usually limited by the extent of a controlled vocabulary, is often journal specific, can be affected by indexing policies, and cannot necessarily accommodate new developments due to the time it takes to incorporate new concepts to a controlled vocabulary.

## 3   Results

We start by showing, in Figure 5, the publication trends for all three fields with regard to volume. The purpose of this figure is to compare publication trends with the ones we will derive for the trends of cognitive extent. Physics generally sustained an exponential growth in the number of articles throughout the 20th century and continuing into the 21st, with a doubling time of 16 years. There is a marked dip coinciding with World War II (WWII), when many physicists apparently engaged in activities that did not lead to published papers (Perc, 2013). Interestingly, astronomy saw no such decline in that period. Unlike physics, the exponential rise in publications in astronomy started only after WWII, and now has a doubling time of 12 y. Biomedicine also exhibits an exponential rise in publication volume, but with somewhat longer doubling time of 19 y.

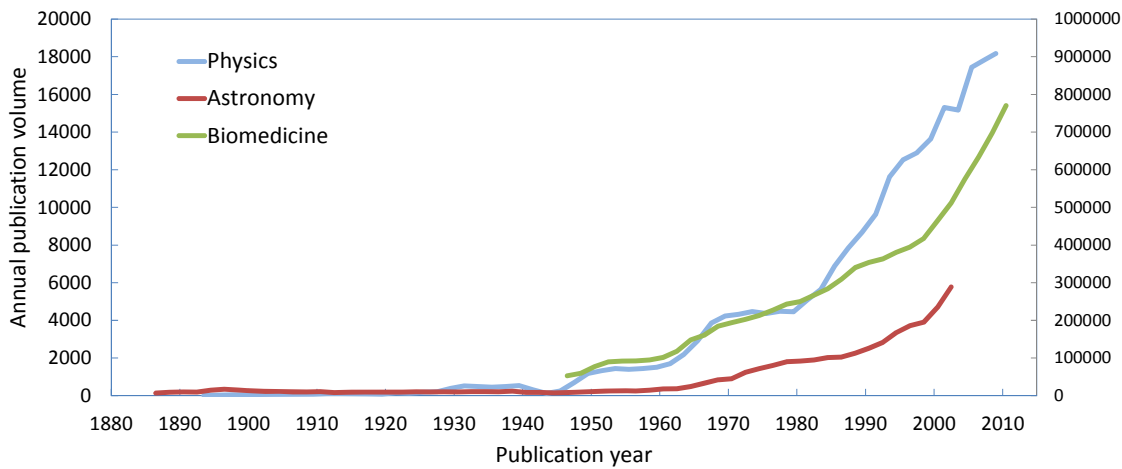

**Fig. 5. Number of articles published annually in physics, astronomy, and biomedicine.** All fields exhibit an exponential growth in the number of publications in recent decades. Data points are averaged in 2-year intervals. Physics experienced a five-fold drop during World War II. Left axis applies to physics and astronomy, while right axis applies to biomedicine.

Figure 6 shows the results of the application of the proposed method to obtain the evolution of the extents of cognitive domains. Each data point in Figure 6 has been derived independently from every other point, using only the unit quota of articles from that time period, and thus having no "history" of what preceded. Nevertheless, all fields show general expansion, reflecting the cumulative nature of science. However, unlike the hundred- to a thousand-fold rise in the publication volume, the growth of cognitive extent is within a factor of few. Furthermore, the trends are often different compared to the changes in the publication volume, even when the latter are presented in logarithmic form (inset panels in Figure 6). For example, the fivefold drop in the publication rate in physics during WWII led to no reduction in its cognitive extent, i.e., the breadth of topics was maintained even with fewer publications. Another example of a disconnect between the publication trends and intellectual content is the stasis in publication volume in physics in the 1970s, which was nevertheless the period of its fastest cognitive growth.

In recent times, where the volume of publishing allows measurements of the cognitive content based on multiple independent article quotas within the same year, we obtain values with very small dispersion (<1.5%), attesting to the robustness of the measure. Thus, if we were presented with a batch of articles from an unknown time period, the measurement of its cognitive extent would typically allow the determination of the publication period to within a decade.

While the overall trend in Figure 6 is the one of growth, one can identify phases of slower or faster changes. We determine these phases by eye and show in Figure 6. In physics, there is little change in the first few decades of the 20th century (labeled P1 in Fig. 6), which may cause initial surprise, as this was the period of the establishment of quantum-mechanical theory and the theories of relativity. However, it is widely known that most of the founders of modern physics worked and published in Europe, while our dataset includes only the American corpus. Indeed, three of the ten founders of quantum mechanics have never published in Physical Review, while others only began publishing there between 1926 and 1948, the time that coincides with a period of mild growth (P2). Very rapid expansion started in the late 1950s (P3), coincident with the well-known "Sputnik effect" (Geiger, 1997). This rapid phase started to slow in the mid-1980s, possibly related to the sharp decline in US physics PhDs, which reached a minimum in 1983 (Mulvey & Nicholson, 2014). The knowledge domain of physics today is expanding (P4) but at a rate that is slower than at any time in the last 50 years. In astronomy we see a rise at the turn of the 20th century (A1), presumably associated with the establishment of astrophysical techniques that greatly extended its domain. American astronomers, and therefore the articles written in English, were at the forefront of this expansion (Bartusiak, 2009). Ensuing decades saw no great change (A2), with modest expansion in the 1940s and 50s (A3). As in physics, the great expansion started in the late 1950s. While not as rapid as in physics, this phase of cognitive expansion in astronomy continues without much change in pace until today (A3). Finally, in biomedicine, we cover trends after WWII. We identify the 1960s as the period of fastest cognitive expansion (B2), tentatively associated with the final phases of the establishment of molecular biology (Morange, 2000).

We caution the reader that it is not justified to use the derived measures to compare the cognitive extents of different fields. This is because different fields may have inherently different lexical structure in their titles or have more extensive objects of study. In this work all comparisons are made within a given field, either over time, or over different groups of knowledge producers.

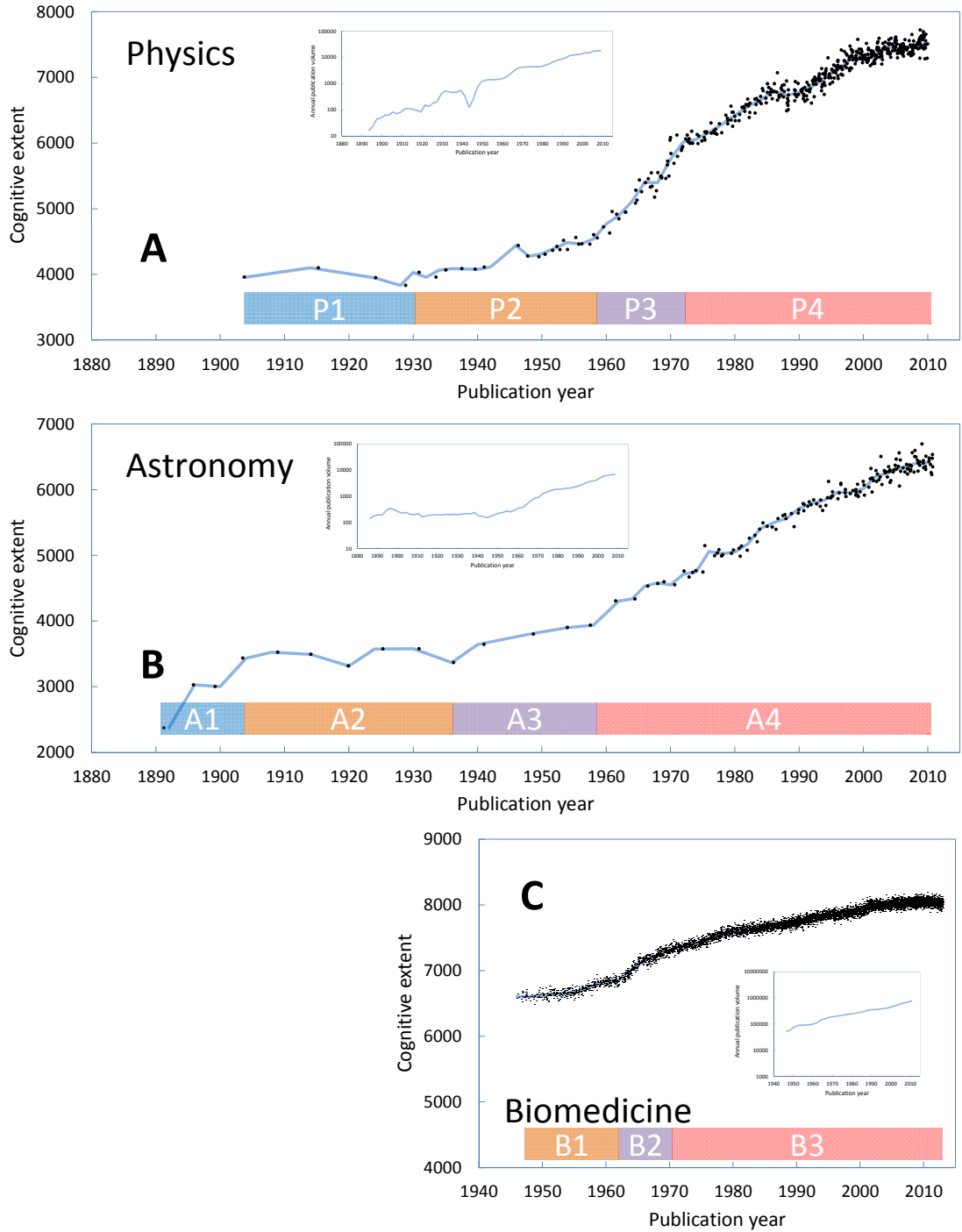

**Fig. 6. Evolution of the sizes of cognitive domains in physics, astronomy, and biomedicine over the last 60-120 years.** Cognitive evolution shows trends that are different from the trends in publication volume (Figure 5). Some general similarities can be seen only after the trends in publication volume are shown in logarithmic form (insets in each panel), but even then, physics is significantly different. Various phases of faster or slower growth have been approximately identified (P1-P4, A1-A4, B1-B3) and are discussed in the text. Blue lines show measurements averaged in 2-year intervals. Statistical uncertainties (one standard deviation) of the average trends in recent times are comparable or smaller to the width of the lines (see also Fig. 3). Each data point is based on approximately 1000-3000 articles.

Figure 7 addresses the research question of how the knowledge produced by teams of different sizes (including single-author "teams") compares in terms of the extent of cognitive areas. Here, all measurements pertain to the most recent time period and each data point is based on the same quota of literature. Remarkably, in both physics and astronomy, the works of single authors, author pairs, and small teams cover the greatest cognitive extent, the same size as that of the entire field. However, beyond several authors, the team size and the cognitive extent of their papers are in inverse correlation. In physics and astronomy, papers from teams comprising of 10 members cover 85% of the maximum extent. The cognitive extent of papers of yet larger teams drops further, such that those with ~60 members cover 60%. Articles produced by 'big science' teams in physics (>100 authors) cover only 35%. Biomedicine follows a similar inverse correlation between cognitive extent and team size, but unlike physics and astronomy the size of the cognitive domain peaks for papers coming from teams having 3-5 members, while it is lower for papers produced by single authors.

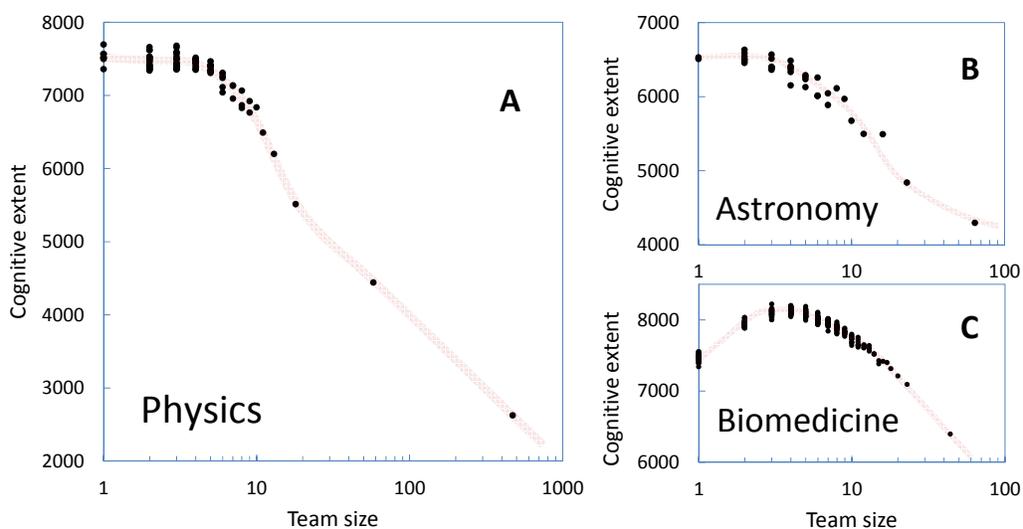

**Fig. 7. Correlations between cognitive extents and the sizes of teams**. In physics and astronomy, papers by single authors, pairs of authors, and small teams cover the largest intellectual territory, the same size as the entire field. Papers from larger teams cover significantly smaller cognitive territory. In biomedicine, papers from teams of 3-5 authors cover the largest domain, but as in astronomy and physics, papers produced by larger teams cover smaller cognitive territory. Trends are derived from the most recent publications (the last five years for physics and astronomy and one year for biomedicine). Each point is derived from the same quota of journal articles.

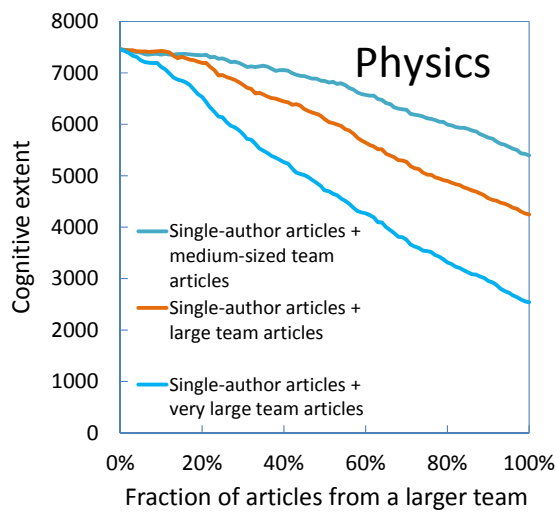

**Fig. 8. Test of the exclusivity of the cognitive domains covered by large teams**. In physics and astronomy larger teams cover smaller cognitive extent than single authors. However this may still be an exclusive territory. Here we test this possibility by replacing a fraction of articles produced by single authors with those from larger teams (binned in three categories of size) and calculate the resulting cognitive extent of the mix. In all cases the extent drops, i.e., the larger teams do not cover exclusive cognitive domains.

While articles produced by larger teams evidently have smaller cognitive extent than smaller teams, this coverage may nevertheless be exclusive, lying beyond the extent of knowledge produced by small teams. The fact that the extent of papers of small teams is similar to that of the entire field suggests that this is most likely not the case, but we still wish to test this possibility directly. We take a random single quota of articles produced by single authors in physics, which of all team sizes have the largest extent (Figure 7), and replace a fraction of those articles with articles produced by larger teams. If the work produced by larger teams covered exclusive cognitive areas, then the addition of such articles to those from single authors would initially lead to the expansion of the cognitive extent. The results of the test are shown in Figure 8, where we replace articles from single authors with those from medium-sized teams (~20 authors), large teams (~60 authors) and very large teams (~500 authors). The cognitive extents (the count of unique concepts) of these mixed quotas immediately start to drop, i.e., the phrases from larger teams only increase the redundancy. We therefore conclude that no part of the cognitive territory is exclusive to large teams.

## 4  Discussion and conclusions

The presented results paint a different picture of the development of scientific fields from the one obtained considering trends purely from publication volumes. While publication rates continue to grow exponentially, the cognitive domains expand on linear scales, in closer agreement with our intuitive notions of the growth of knowledge, and better matching the changes in the breadths of graduate curricula or science textbooks, which have obviously not grown exponentially. The measure of the cognitive extent, and its change, also provide a clearer connection with societal circumstances in which science develops. Our results show that the rapid expansion in physics and astronomy started in the decade following WWII, suggesting that the post-war investments in science had a great effect on the intellectual expansion of these fields. The period of the fastest expansion of biomedicine in the 1960s also coincides with a large increase in funding, and the influx of researchers from other fields that led to major advancements at the intersection with analytical chemistry, physics, and computer science.  The drop in interest, as evidenced by the number of new PhDs, and the funding in physical sciences in the 1970s may have led to somewhat slower rate of expansion in physics, but it did not affect astronomy, which, lacking the applied component, was less affected by declining R&D funding.

Another result of the proposed method is that the science produced by large teams in general covers significantly smaller knowledge domain than the scientific output of small teams, or, in the case of physics and astronomy, even of individuals, despite the fact that single authors now contribute less than a tenth of articles produced. Previous studies have shown that small and large teams have different formation mechanisms (Milojević, 2014), whereby small teams, required to solve most scientific problems, form by a Poisson process characterized by a small number of researchers, while large teams, presumably formed to solve different types of questions, acquire new members on the principle of cumulative advantage. Our results regarding the cognitive extent suggest that the large teams are more specialized in nature and consequently do not encompass the intellectual breadth of the entire discipline. On the other hand, the small teams cover all aspects of knowledge and are thus possibly responsible for maintaining or even expanding the scientific territory. The maximum cognitive extent coverage by works of single authors in physics and astronomy can be explained by the fact that these fields also contain a pure theoretical component. Theoreticians do not require large facilities, tend to work alone, and strive to make a contribution in every research topic of some field. Works of single authors in biomedicine, on the other hand, cover more limited territory, presumably because in this field, dominated by experimental results, individuals are significantly limited in terms of the scientific questions they can address without having a lab, and therefore a team, even a small one.

Another potential driver of the declining cognitive extent with the increasing team size may be the number of unique authors that are included in article quotas, i.e., the possibility that the cognitive quota scales with the number of unique authors. If large-team article quotas had significantly fewer unique authors, it could provide some explanation for the observed trends. We test this for the case of astronomy and find that the largest-team quota actually has 10 times as many unique authors as the single-author quota, so taken at face value this does not provide an explanation. The results are different, however, if only the first (corresponding) authors are considered. The number of unique first authors peaks for teams of two and three authors. So it is still the case that there are fewer unique first authors in single-author quotas, even though they have the same cognitive extent as the quotas from articles from team sizes of two and three. We will explore the connection between cognitive extent and unique authors and their individual contributions in subsequent work.

The results of this study indicate that "little science" still has an important role in expanding the domains of knowledge and maintaining its heterogeneity and should therefore not be neglected on account of big-team projects. The application of the method for quantifying the cognitive extent in a body of literature presented in this paper can have numerous applications in research policy and research evaluation, such as exploring the development of science in different geographical or institutional entities, by gender, or some other author characteristic. The results of such studies will have an advantage over often distorted citation measures, and will not be affected by mere production volumes. The proposed approach holds promise for a better understanding of many questions regarding the 'science of science.'

## Acknowledgments

I gratefully acknowledge the discussions and the encouragement from Kevin Boyack and Richard Klavans. I thank Colleen Martin for copyediting. I also thank anonymous referees for their constructive comments.